\documentclass[aip,sd,amsmath,amssymb,preprint,]{revtex4-1}
\usepackage{graphicx}
\usepackage{epstopdf, epsfig}

\usepackage[normalem]{ulem}
\usepackage[shortlabels]{enumitem}
\usepackage{CJK}
\usepackage{amssymb,euscript,latexsym,amsmath}
\usepackage{graphicx}
\usepackage{color}
\usepackage{epstopdf}
\usepackage{setspace}
\usepackage{subfigure}
\usepackage{chngpage}
\usepackage{textgreek}

\usepackage{tikz}

\newcommand{\tikzcircle}[2][black,fill=black]{\tikz[baseline=-0.5ex]\draw[#1,radius=#2] (0,0) circle ;}
\newcommand{\tikzdiamond}[1][fill=black]{\tikz [x=1.2ex,y=1.5ex,line width=.1ex,line join=round, yshift=0ex] \draw  [#1]  (0,.5) -- (.5,1) -- (1,.5) -- (.5,0) -- (0,.5) -- cycle;}
\newcommand{\tikzbox}[1][fill=black]{\tikz [x=1.2ex,y=1.2ex,line width=.1ex,line join=round, yshift=0ex] \draw  [#1]  (0,0) -- (0,1) -- (1,1) -- (1,0)  -- cycle;}

\begin{document}


\begin{flushleft}
{\Large
\textbf\newline{Bistability in the rotational motion of rigid and flexible flyers}
}
\newline
\\
Yangyang Huang$^1$, Leif Ristroph$^2$, Mitul Luhar$^1$,
 and Eva Kanso$^1$\textsuperscript{*}

\bigskip
1. Aerospace and Mechanical Engineering, University of Southern California, Los Angeles, California, USA\\
2. Courant Institute of Mathematical Sciences, New York University, New York, New York 10012, USA
\bigskip

* kanso@usc.edu

\end{flushleft}

\section*{abstract}
We explore the rotational stability of hovering flight. Our model is motivated by an experimental pyramid-shaped object (\citet{Weathers2010, Liu2012}) 
and a computational $\wedge$-shaped analog (\citet{Huang2015a, Huang2016}) hovering passively in oscillating airflows;
both systems have been shown to maintain rotational balance during free flight. Here, we attach the $\wedge$-shaped flyer at its apex, allowing it to rotate freely akin to a pendulum.
We find that the flyer exhibits stable concave-down ($\wedge$) and concave-up ($\vee$) behavior. 
Importantly, the down and up configurations are bistable and co-exist for a range of background flow properties.
We explain the aerodynamic origin of this bistability and compare it to the inertia-induced stability of an inverted pendulum oscillating at its base.
We then allow the flyer to flap passively by introducing a rotational spring at its apex.  For stiff springs, flexibility diminishes upward stability but as stiffness decreases, a new transition to upward stability is induced by flapping. We conclude by commenting on the implications of these findings for biological and man-made aircraft.


\section{\label{sec:level1}Introduction}
Stability is as essential to flight as lift itself. Flyers, living and nonliving, are often faced with perturbations in their environment. 
After a perturbation, a stable flyer returns to its previous orientation passively. An unstable one requires active control. 
The issues of stability and control were indispensable to the development of man-made aircrafts~\citep{Wright1906} and are 
pertinent to both the origin of animal flight and the subsequent evolution of flying lineages~\citep{Vogel2009}.

The intrinsic stability of flying organisms varies across species. Most birds and insects sacrifice intrinsic stability for gains in maneuverability and performance~\citep{Vogel2009, Huang2015a}. This trade-off is enabled by sensory feedback and neuromuscular control mechanisms. 
To identify the sensory circuits and control strategies employed by insects, several approaches have been used.  This includes pioneering behavioral experiments as well as anatomical and aerodynamic studies~\citep{Fry2003,Taylor2007,Ristroph2010,Sun2014}. 

Insects have also been a source of inspiration for building miniature flying machines; see, for example, \citet{Ma2013} and references therein. 
Most designs imitate the  flapping motions of insect wings. The aerodynamics of these flapping motions have been clarified by numerous experimental and computational models; see, for example,~\citet{Ellington1996, Dickinson1999,Sane2003,Wang2004,Wang2005} and references therein. However, stabilization and control of such biomimetic machines remains a challenge; it requires fast responses to unsteady aerodynamics at small length scales.  It is therefore advantageous to invent new engineering designs that are intrinsically stable. To this end, \citet{Ristroph2014} proposed a {jellyfish-inspired} machine that required no feedback control to achieve stable hovering and vertical flight.  The aerodynamic principles underlying this stable hovering are fundamentally linked to a previous experimental model by the same research team where a pyramid-shaped object pointing upward was shown to hover and maintain balance passively, without internal actuation, in vertically-oscillating airflows of zero mean~(\citet{Childress2006, Weathers2010,Liu2012}). Building on these efforts,~\citet{Huang2015a,Huang2016} analyzed the aerodynamics and stability of the pyramid-shaped flyer using a  two-dimensional computational model based on the inviscid vortex sheet method.  
\citet{Fang2017} applied a similar approach to examine the stability of the {jellyfish}-inspired hovering machine. Details of the vortex sheet method can be found in~\citet{Krasny1986, Nitsche1994,Jones2003,Jones2005,Shukla2007,Alben2009}. 

Stable hovering in oscillating flows of zero-mean is enabled by the pyramid's geometric asymmetry and the unsteady vortex structures shed from its outer edges. \citet{Liu2012} combined experimental observations with a quasi-steady force theory to  estimate the effect of this asymmetry without ever solving for the coupled fluid-flyer interactions. They reported that, contrary to intuition, pyramids with higher center of mass are more stable. 
Coupled fluid-flyer interactions were computed by~\citet{Huang2015a,Huang2016} in the context of a two-dimensional $\wedge$-shaped flyer free to undergo translational and rotational motions in oscillating flows. These computational studies provided valuable insight into the background flow conditions necessary for hovering and into hovering stability. In particular, a transition from stable to unstable, yet more maneuverable, hovering was reported as a function of the flyer's opening angle and background flow acceleration.

As an extension of the research reported in~\citet{Huang2015a,Huang2015b,Huang2016}, we consider here the rotational stability of a heavy $\wedge$-shape flyer that is attached at its apex, but free to rotate, in a vertically-oscillating background flow.
As in a simple pendulum, the $\wedge$-configuration is stable and the $\vee$-configuration is unstable in the absence of flow oscillations.  We first consider rigid flyers and examine the stability of these two configurations in oscillating flows. We find that aerodynamics stabilizes the upward $\vee$-configuration for a range of background flow parameters, namely, amplitude and frequency of oscillations. We compare these parameters to those required to stabilize a `dry' pendulum in the upward configuration by fast vertical oscillations at its base, i.e., the classical inverted pendulum. We find that aerodynamics can induce upward stability at lower oscillation frequency and amplitude.
Importantly, the upward configuration can be stable even under perturbations as large as $\pi/2$.  We explain the aerodynamic origin of this bistability about the downward and upward configurations by analyzing in detail the aerodynamic forces and torques acting on the flyer. To do so, we employ the vortex sheet model and we develop a quasi-steady point force model that takes into account the geometry of the flyer. Lastly, we introduce a rotational spring at the apex of the flyer and allow it to flap passively under background flow oscillations. We find that the flapping frequency is always slaved to the frequency of the background flow.  For the parameter ranges considered here, the intrinsic frequency of the flyer does not play a role. Elasticity diminishes upward stability in stiff flyers.  However, with further decreases in stiffness, a new transition to upward stability is observed. This transition is induced by large-amplitude flapping motion of the flyer.

\section{\label{sec:level2}Problem formulation}

\begin{figure*}
\centering
\includegraphics[width=\linewidth]{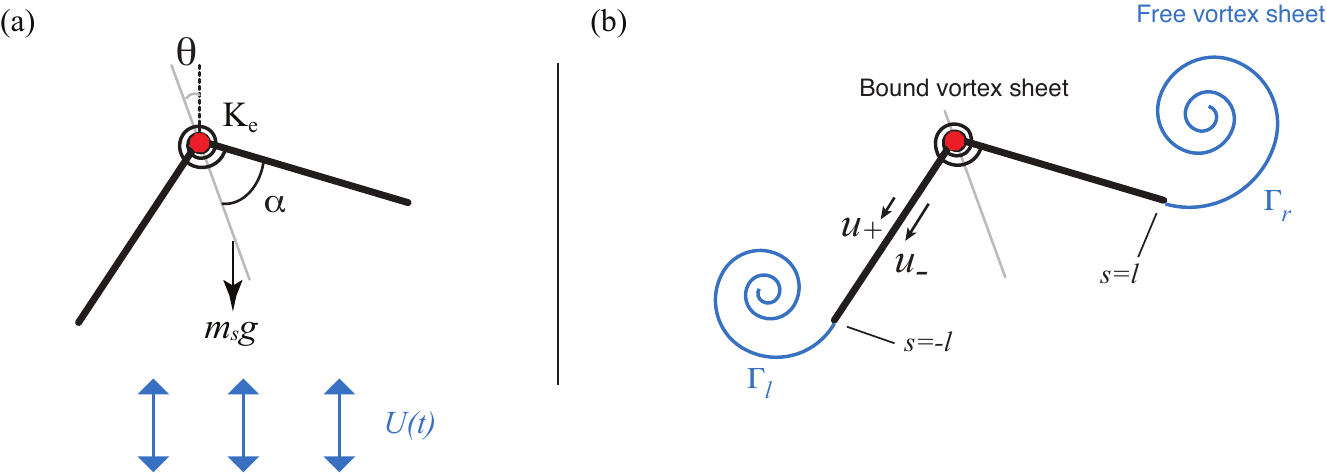}
\caption{\footnotesize (a) Schematic of the two-dimensional $\wedge$-shaped flyer in oscillatory fluid.
(b) Depiction of the vortex sheet model used for calculating the aerodynamic forces and torques on the flyer.
}
\label{fig:schematics}
\end{figure*}
 
 The flyer consists of two flat `wings' connected rigidly at their apex to form a $\wedge$-flyer, as shown in figure~\ref{fig:schematics}(a).
The opening angle of the flyer is $2\alpha$. The wings are made of rigid plates of homogeneous density  $\rho_s$, length $l$, and thickness $e$ that is small relative to $l$.  The mass per unit depth of each wing is given by $m_s = \rho_s l e$.
The flyer is suspended at its apex $O$ but free to rotate about $O$ at an angle $\theta$ measured counterclockwise from the vertically-up direction. The flyer is placed in a background flow of density $\rho_f$ oscillating vertically at a velocity $U(t) = \pi fA \sin(2\pi ft)$ with zero mean. Here, $f$ is the oscillation frequency and $A$ is the peak-to-peak amplitude.

The equation governing the rotational motion of the flyer is obtained from the conservation of angular momentum about point $O$ of the two-wing system subject to gravitational and aerodynamic effects,
\begin{equation}
\frac{2}{3}m_sl^2\ddot{\theta} = -(m_s-m_f) gl\cos\alpha\sin\theta + {T}_1 + {T}_2.
\label{eq:theta}
\end{equation}
Here,  $g$ is the gravitational constant, $m_f=\rho_f l e$ is the mass of displaced fluid, and $(m_s - m_f)g$ is the net weight of each wing counteracted by the buoyancy effects. 
The aerodynamic torques on the left and right wings respectively are denoted by $T_1$ and $T_2$, resulting in a total aerodynamic torque ${T} = {T}_1 + {T}_2$ about the flyer's  point of suspension $O$. If these torques were zero,~\eqref{eq:theta} reduces to the equation $\ddot{\theta} = -(g/l_{\rm p}) \sin\theta$ governing the rotational motion of a simple pendulum of  length $l_{\rm p} =  (2m_s / 3(m_s - m_f)) (l /\cos\alpha)$.

For the flexible flyers, we introduce elasticity into the model in the form of a torsional spring of stiffness $K_e$ placed at the base point connecting the two rigid wings. 
For this case, in addition to the rotational dynamics in~\eqref{eq:theta}, the shape of the flyer, represented by the half-opening angle $\alpha$, changes in time. The equation of motion governing the shape evolution is obtained by balancing the angular momentum for each wing separately and subtracting the resulting two equations.  This yields
\begin{equation}
\begin{split}
\frac{2}{3}m_s l^2\ddot{\alpha} &= -(m_s-m_f) gl\sin\alpha\cos\theta - ({T}_1 - {T}_2) - 4K_e(\alpha-\alpha_r).
\label{eq:alpha}
\end{split}
\end{equation}
Here, $\alpha_r$ is the rest half-angle of the torsional spring.

To make the equations of motion~\eqref{eq:theta} and~\eqref{eq:alpha} dimensionless, we scale length by $l$, time by $1/f$, and mass by the wing's added mass $\rho_f l^2$. The number of independent parameters is then reduced to five dimensionless quantities: the amplitude $\beta$ and acceleration $\kappa$ of the background flow and the mass
$m$, rest angle $\alpha_r$, and stiffness $k_e$ of the flyer,
\begin{equation}
\begin{split}
\label{eq:dimensionless}
\beta = \frac{A}{l}, \quad \kappa = \frac{2m_sAf^2}{3(m_s - m_f)g}, \quad  m = \frac{2m_s}{3\rho_f l^2}, \quad   \alpha_r, \quad k_e = \frac{4K_e}{\rho_f l^4 f^2}.
\end{split}
\end{equation}
For the rigid flyer,  $k_e =\infty$ and $\alpha=\alpha_r$ for all time.
Dimensionless counterparts to~\eqref{eq:theta} and~\eqref{eq:alpha} can be written as
\begin{equation}
\begin{split}
m\ddot{\theta} &= -\frac{m\beta}{\kappa}\cos\alpha\sin\theta + (T_1 + T_2), \\
m\ddot{\alpha} &= -\frac{m\beta}{\kappa}\sin\alpha\cos\theta - (T_1 - T_2) -k_e(\alpha-\alpha_r).
\label{eq:eom}
\end{split}
\end{equation}
Here, the aerodynamic torques $T_1$ and $T_2$ are considered to be dimensionless. The dimensionless background flow is given by $U(t) = \pi \beta \sin(2\pi t)$.

\section{The vortex sheet method}
We apply an inviscid vortex sheet model to calculate the aerodynamic forces and torques exerted on the flyer by the surrounding fluid. A detailed description of the vortex sheet method can be found in~\cite{Huang2016} and references therein. Here, we give a brief outline of the method. In this treatment, the wing system is modeled as a bound vortex sheet of zero thickness and the vorticity shed at each edge is represented as a free vortex sheet, as shown in figure~\ref{fig:schematics}(b).  Vorticity is distributed along the free and bound vortex sheets with sheet strength $\gamma(s,t)$, as a function of the arc length $s$ and time $t$. We define the total circulation of the left and right vortex sheets as $\Gamma_l = \int_{s_l}\gamma(s,t)ds$ and $\Gamma_r = \int_{s_r}\gamma(s,t)ds$ respectively. Here, $s_l$ and $s_r$ are used to denote the arc-lengths along the left and right vortex sheets. The distribution of the bound sheet strength at each time step is solved by satisfying the normal boundary conditions on the wings and Kevin's circulation theorem. The Kutta condition gives the shedding rates at the two outer edges as 
\begin{equation}
\frac{d\Gamma_l}{dt}|_{s_b=-l} = -\frac{1}{2}(u_-^2 - u_+^2)|_{s_b=-l}, \quad \frac{d\Gamma_l}{dt}|_{s_b=l} = \frac{1}{2}(u_-^2 - u_+^2)|_{s_b=l},
\end{equation}
where $s_b$ is the arc length along the bound vortex sheet ($s_b = -l$ and $s_b = l$ denote the arc lengths of the left and right edges separately) and $u_\pm$ are the slip velocities above and below the flat wings, namely the tangential velocity difference between the fluid and the wing.

Once the vorticity distribution is computed, the pressure difference across the wings can be obtained from Euler's equation. To this end, we get
\begin{equation}
[p]_+^-(s_b,t) = p_-(s_b,t) - p_+(s_b,t) = -\frac{d\Gamma(s_b,t)}{dt} - \frac{1}{2}(u_-^2-u_+^2),
\end{equation}
where $\Gamma(s_b,t) = \Gamma_l + \int_{-l}^{s_b} \gamma(s,t)ds$.
The fluid force is due to pressure only; the force and torque acting on each wing with respect to the attachment point $O$ are given by
\begin{equation}
\begin{split}
 &F_x  =   \int_{-l}^{l}[p]_+^-n_x ds, \qquad F_y = \int_{-l}^{l}[p]_+^- n_y ds, \\
& T = \int_{-l}^{l}[p]_+^- \bigl((x_b-x_o)n_y - (y_b-y_o)n_x\bigr) ds.
\label{eq:aero}
\end{split}
\end{equation}
Here, $n_x$ and $n_y$ are the $x$- and $y$-components of the unit vector normal to the wings, $(x_b,y_b)$ is the position of the bound vortex sheet along the wings, and $(x_o, y_o)$ is the fixed position of the attachment point. Both $(x_b,y_b)$ and $(n_x,n_y)$ are functions of arc-length.

To emulate the effect of fluid viscosity, we introduce a dimensionless time parameter $\tau_{\rm diss}$, such that the point vortices shed at time $t-\tau_{\rm diss}$ are manually removed from the fluid at time $t$. Larger $\tau_{\rm diss}$ indicates smaller fluid viscosity. 
In this paper, we choose $\tau_{\rm diss} = 0.6$ to be in the order of the oscillation period $\tau = 1$, as explained in~\citet{Huang2015a,Huang2016}. We expect the results to be qualitatively similar for variations in $\tau_{\rm diss}$ between $0.6$ and $1$;  see~\citet{Huang2015a}.

\begin{figure*}
\centering
\includegraphics[scale = 1]{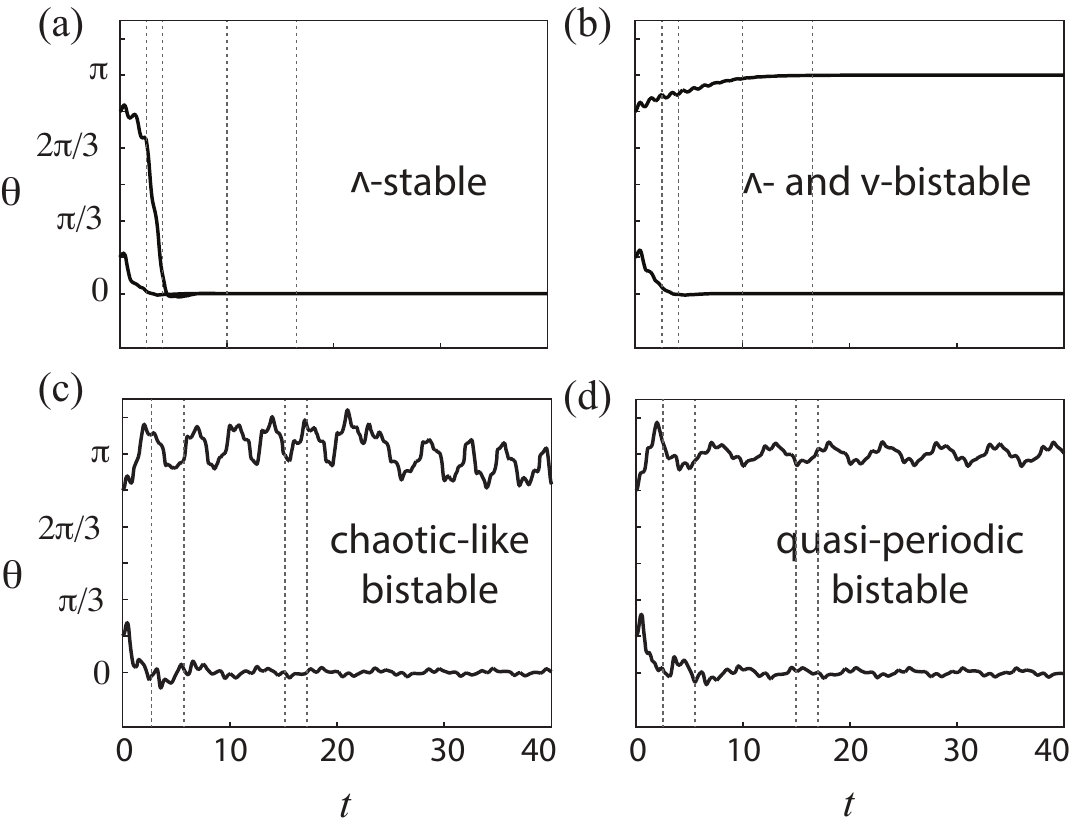}
\caption{\footnotesize 
Rotational behavior of a rigid flyer ($m= 1$, $\alpha = \pi/6$) in oscillatory flows. Flow parameters are (a) $\beta= 0.6$, $\kappa = 0.15$, (b) $\beta= 0.6$, $\kappa = 0.30$, (c) $\beta= 1.2$, $\kappa = 0.15$
 and (d) $\beta= 1.2$, $\kappa = 0.30$.   
Initial perturbations are set to $\theta(0)=\pi/6$ and $\theta(0) = 5\pi/6$. Snapshots of the flyer's wake at the time instants highlighted by vertical dashed lines are shown in figure~\ref{fig:rigidshedding}.}
\label{fig:rigidflyer}
\end{figure*}

\begin{figure*}
\centering
\includegraphics[width=\linewidth]{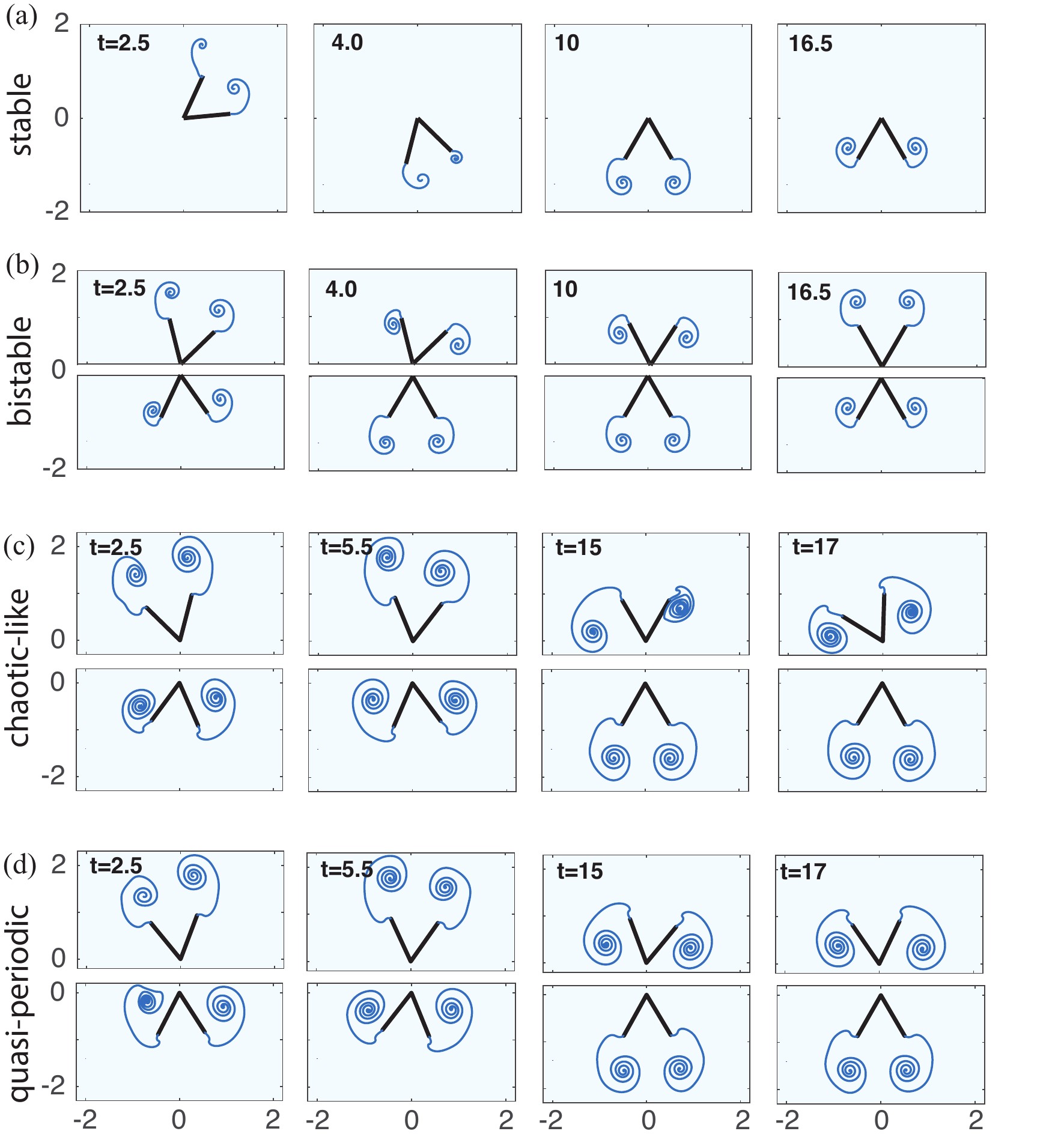}
\caption{\footnotesize Snapshots of the flyers and their  wakes at the time instants highlighted by vertical dashed lines in figure~\ref{fig:rigidflyer}. 
}
\label{fig:rigidshedding}
\end{figure*}

\section{Results: rigid flyers}

The concave-down ($\wedge$)  and concave-up ($\vee$)  configurations of the flyer are equilibrium solutions of~\eqref{eq:eom}. This result follows directly from symmetry about the vertical direction. In the absence of flow oscillations, as in a simple pendulum, the $\wedge$-configuration is stable and the $\vee$-configuration is unstable.  Here, we examine the stability of these two configurations in oscillating flows by solving the nonlinear system of equations for the coupled fluid-flyer model. For concreteness, we consider perturbations of the flyer's initial orientation $\theta(0)$ while keeping $\dot{\theta}(0) = 0$. For the elastic flyer {discussed in Section \ref{sec:elastic}}, we additionally set $\alpha(0) - \alpha_r = \dot{\alpha}(0) = 0$.

 \subsection{\label{sec:level4}Bistable behavior}

We impose non-zero initial perturbations $\theta(0)$ and we solve~\eqref{eq:eom}, coupled to the vortex sheet model, for each initial perturbation. Figure~\ref{fig:rigidflyer} shows the rotational motion $\theta(t)$ of a flyer of  mass $m=1$ and half-opening angle $\alpha = \pi/6$ for four sets of flow parameters  $(\beta, \kappa) = (0.6, 0.15)$,  $( 0.6, 0.3)$,  $(1.2, 0.15)$, and  $(1.2, 0.3)$. Figure~\ref{fig:rigidshedding} shows snapshots of the flyer and its unsteady wake for these four cases. Two distinct nonlinear behaviors are observed: stable behavior where the flyer gravitates to the concave-down $\wedge$-configuration for all initial perturbations {(Figure~\ref{fig:rigidflyer}(a))} and bistable behavior where the flyer tends to either the concave-down $\wedge$-  or concave-up $\vee$-configuration depending on the initial perturbation {(Figure~\ref{fig:rigidflyer}(b))}. We further distinguish three types of bistable behavior: asymptotically stable behavior where $\theta$ converges  to either $0$ or $\pi$ (Figure~\ref{fig:rigidflyer}(b)),  bounded `chaotic-like' oscillations about $0$ or $\pi$ (Figure~\ref{fig:rigidflyer}(c)), and  `quasi-periodic' oscillations about $0$ or $\pi$ (Figure~\ref{fig:rigidflyer}(d)). Similar bounded oscillations were observed in the stable behavior about the concave-down $\wedge$-configuration, the time trajectories of which are omitted for brevity. 

Stabilization of the flyer in the concave-up $\vee$-configuration is fundamentally due to unsteady aerodynamics. Snapshots of the flyers and their unsteady wakes  are shown in figure~\ref{fig:rigidshedding}. 
The flyer is subject to gravitational and aerodynamic forces only. The torque $-m\beta\kappa^{-1} \cos\alpha \sin \theta$ induced by the gravitational force tends to align the flyer  with $\theta =0$ for all orientations. Thus, it has a destabilizing effect on the concave-up $\vee$-configuration. 
Later in this paper we analyze the aerodynamic forces $F_x$ and $F_y$ and torque $T$ acting on the flyer and explain the aerodynamic origin of the bistable behavior. First, we map the flyer's stable and bistable behavior onto the two-dimensional parameter space ($\beta$, $\kappa$) of flow amplitudes and accelerations.

  \begin{table}
    \begin{center}
        \begin{tabular}{ c  c c c c | c c c c c } 
            \multicolumn{5}{c|}{Figure~\ref{fig:paramspace}(a)} & \multicolumn{5}{c}{Figure~\ref{fig:paramspace}(b)}\\\hline
         $\alpha$ & $ \pi/12$  & $ \pi/6$  & $ \pi/4$ & & & $m$  & $0.5$  & $1$ &  $ 2$ \\ \hline
          $a$ &  $-0.63$ 
           & $-0.45 $ & $-0.37 $  & & & $a$ & $-0.52$ & $-0.57 $ &  $-0.55 $\\[1ex] 
           $b$ & $ 0.23$ & $ 0.17$ & $ 0.16$  & & & $b$  &$ 0.18$ & $0.22$ & $0.23$\\ 
        \hline
        \end{tabular}
    \caption{Transition from stable to bistable behavior first occurs at $\beta / \kappa^a > b$ where $a$ and $b$ are obtained from linearly fitting the lower boundary of the bistable (green) region in figure~\ref{fig:paramspace}. 
    }
    \label{table2}
    \end{center}
\end{table}

\begin{figure*}
\centering
\includegraphics[width=0.85\linewidth]{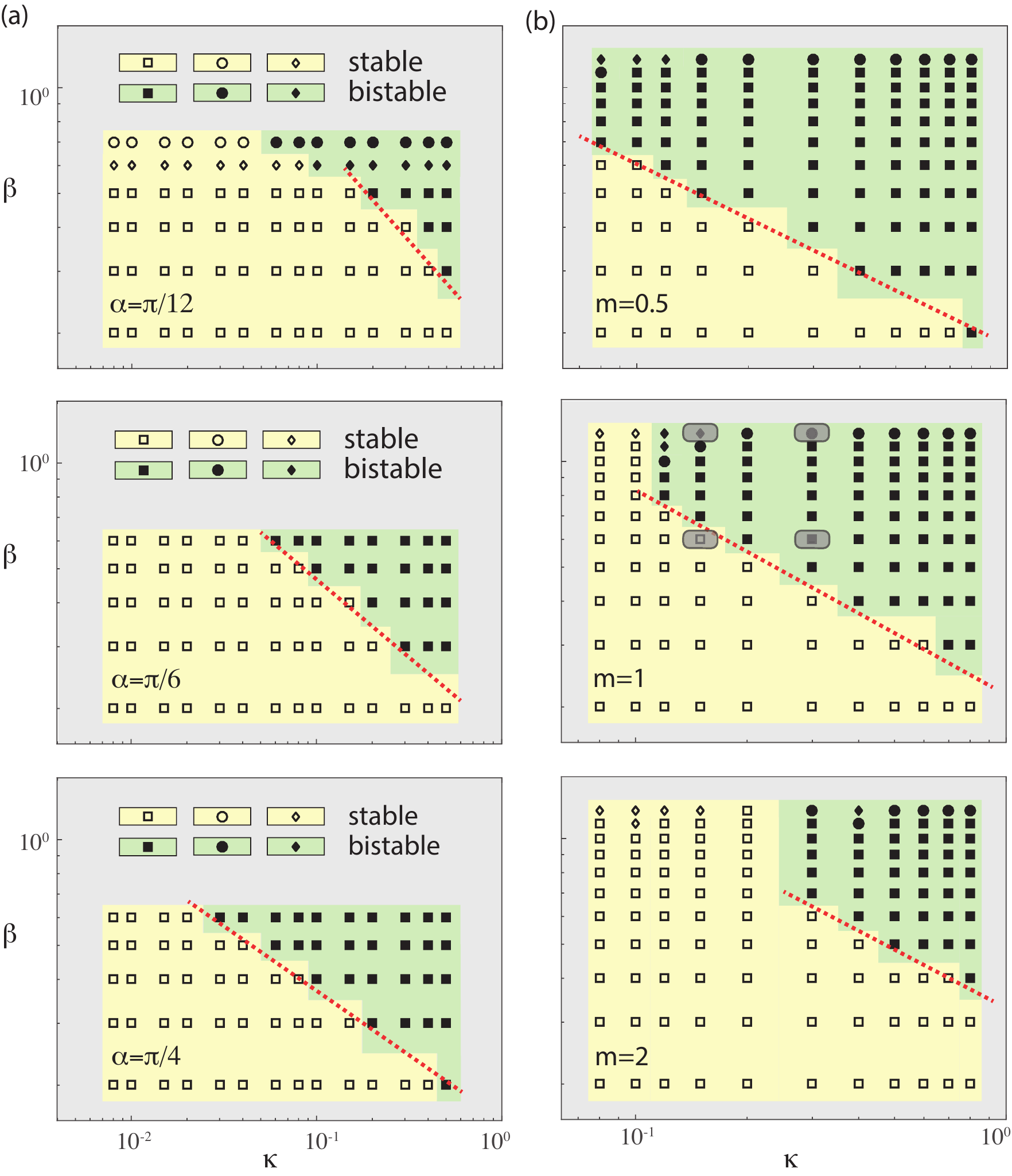}
\caption{\footnotesize  Stable and bistable behavior of rigid flyers mapped onto  the ($\kappa$,$\beta$) for (a) $\alpha=\pi/12, \pi/6$, and $\pi/4$ and $m = 0.074$ and (b) $\alpha = \pi/6$ and  $m=0.5,1,2$. Grey boxes are used to highlight the four sets of parameters used in figure~\ref{fig:rigidflyer}.
}
\label{fig:paramspace}
\end{figure*}

Figure~\ref{fig:paramspace} shows  the ($\beta, \kappa$)-space in log-log scale for six distinct flyers:  three flyers of  increasing opening angle $\alpha= \pi/12, \pi/6$, and $\pi/4$ and same mass $m=0.074$ (figure~\ref{fig:paramspace}(a)), and three  flyers of the same angle $\alpha = \pi/6$ and increasing mass $m=0.5$, $1$, and $2$ (figure~\ref{fig:paramspace}(b)). 
Stable behavior about the concave-down $\wedge$-configuration is represented by the open symbols 
`\tikzbox[fill=white]' `\tikzdiamond[fill=white]'and `\tikzcircle[fill=white]{2.25pt}', corresponding to asymptotically stable behavior ($\theta \to 0$), bounded chaotic-like and periodic oscillations about $\theta = 0$, respectively. The filled symbols `\tikzbox[fill=black]' `\tikzdiamond[fill=black]'and `\tikzcircle[fill=black]{2.5pt}' are used to denote bistable behavior. 
The best-fit line for the points at which the transition from asymptotically-stable  to asymptotically-bistable behavior is first observed is highlighted by a dashed red line, with bistable behavior observed for values of $\beta$ and $\kappa$ values that satisfy
\begin{equation}
\label{eq:bistableflyer}
\beta / \kappa^a > b. 
\end{equation}
The slope $a$ of the transition line and threshold value $b$ above which the transition occurs depend on the flyer's shape $\alpha$ and mass $m$, as detailed in table 1. The slope $a$ increases as $\alpha$ increases but is relatively insensitive to changes in mass.
Meanwhile, 
the threshold $b$ decreases with $\alpha$ but increases with $m$. Taken together, these results indicate that wider flyers, which amplify the aerodynamic torque, tend to transition to bistable behavior at lower values of flow amplitude $\beta$ and acceleration $\kappa$ than narrower flyers. They also indicate that heavier flyers require larger values of $\beta$ and $\kappa$ to make this transition. 

An estimate of the transition from stable to bistable behavior can be obtained by noting that upward stability occurs when the aerodynamic torque balances the gravitational torque.   Considering a quasi-steady drag formulation, the aerodynamic force in dimensional form is proportional to $\rho_f l (A f)^2 $ and the aerodynamic torque  to $\rho_f  l^2 (A f)^2$.   In dimensionless form, one has $T \sim \rho_f l^2 (A f)^2 / \rho_f l^4 f^2 = \beta^2$.
From~\eqref{eq:eom}, the gravitational torque scales as $m\beta \kappa^{-1}\cos\alpha$. Thus, the  ratio of aerodynamic to gravitational torque is given by  $\beta/\kappa^{-1} m\cos\alpha$, yielding $\beta/\kappa^{-1} \geq m\cos\alpha$ for upward stability.  
The threshold to bistability increases with $m$ and decreases as $\alpha$ increases from $0$ to $\pi/2$, which is consistent with the numerical results in figure~\ref{fig:paramspace} and table 1. However, a direct comparison of the condition $\beta/\kappa^{-1} \geq m\cos\alpha$ obtained from such scaling argument with equation \eqref{eq:bistableflyer} implies that $a=-1$ whereas the values listed in table 1 based on the vortex sheet method lie within $-1<a<0$.  This discrepancy indicates that the simple scaling argument based on quasi-steady drag does not quantitatively capture the unsteady flows and associated aerodynamic torques. {A more complete quasi-steady model that describes the aerodynamic origin of the observed bistable behavior is presented in Section~\ref{sec:quasi-steady}.}

\subsection{Comparison to the inverted pendulum}

The bistable behavior observed here is reminiscent to the behavior of a classic pendulum undergoing rapid vertical oscillations about its point of suspension, with negligible aerodynamic forces. A classic pendulum of length  $l_{\rm p} =  (2m_s / 3(m_s - m_f)) (l /\cos\alpha)$ equivalent to the submerged flyer can be stabilized about the inverted (vertically-up) configuration by an inertia-induced torque provided that the frequency $f_{\rm p}$ and amplitude $A_{\rm p}$ of the base oscillations satisfy $A_{\rm p}^2 (2\pi f_{\rm p})^2 > 2gl_{\rm p}$~(\citet[equation (7)]{Butikov2001}). 
The inertia-induced torque responsible for this bistability can be best explained in a non-inertial frame of reference that is oscillating with the base point of the pendulum. The acceleration of this frame induces an inertial torque that must be added to the torque of the gravitational force. Such torque is absent in the flyer equations because the flyer's base point is fixed.  To compare the classic pendulum to the flyer, we rewrite the condition $A_{\rm p}^2 (2\pi f_{\rm p})^2 > 2gl_{\rm p}$ for inertia-induced bistability   in terms of the dimensionless amplitude $\beta = A_{\rm p}/l$ and acceleration $\kappa =   (2m_s/3(m_s-m_f))A_{\rm p}f_{\rm p}^2 / g$ defined according to~\eqref{eq:dimensionless}; we obtain
\begin{equation}
\label{eq:bistablependulum}
\beta /\kappa^{-1} >  \dfrac{1}{2\pi^2\cos\alpha}\left(\dfrac{2m_s}{3(m_s-m_f)}\right)^2 .
\end{equation}
Comparing~\eqref{eq:bistableflyer} and~\eqref{eq:bistablependulum},  $a$ is always equal to $-1$ for the classic pendulum,  reinforcing that $a$ of the flyer is affected by the flyer's shape due to aerodynamics. Meanwhile, the threshold $b$ for the transition to upward stability depends  on both mass and shape, but unlike the trend observed in table 1 for the flyer, $b$ for the inverted pendulum increases as $\alpha$ increases from 0 to $\pi/2$. 

For a quantitative comparison, consider the flyer with $\alpha = \pi/6$ and $m = 1$ (middle panel of figure~\ref{fig:paramspace}(b)). The aerodynamic-induced transition  occurs for $\beta /\kappa^{-0.57}  > 0.22$. If we vary the mass ratio $m_s/m_f$ from 1.2 to 4, the dimensionless quantity $(2m_s/3(m_s-m_f))$ decreases from $4$ to $8/9$ and the threshold value $b$ for the inertia-induced transition decreases by  an order of magnitude from $0.23$ to $0.05$.
At $m_s/m_f = 1.52$, the inertia- and aerodynamic-induced transitions have the same value $b = 0.22$. In this case, for accelerations $0< \kappa< 1$, the flyer transitions to upward stability at  smaller oscillation amplitude $\beta$  than the classic pendulum. By the same token, for a given amplitude $\beta$, this transition requires smaller $\kappa$ and consequently smaller oscillation frequency.
 
\begin{figure*}
\centering
 \includegraphics[scale = 1]{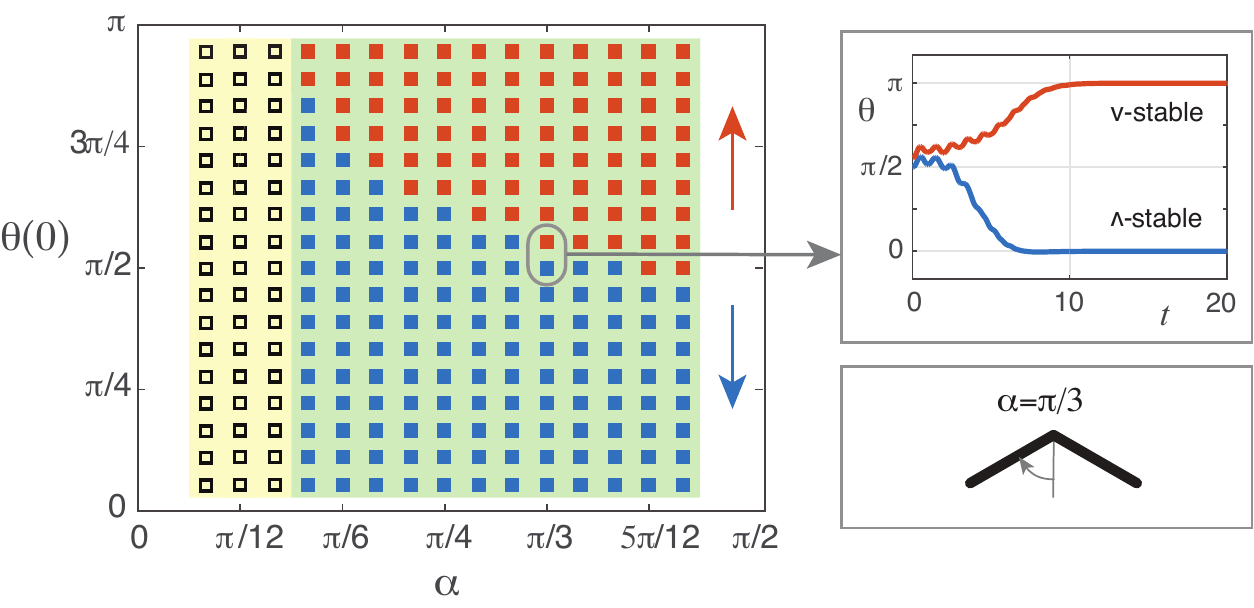}
\caption{\footnotesize Basins of attraction for the (blue) downward $\wedge$- and (red) upward $\vee$-stable configurations for flyers of opening angle $\alpha$ ranging from $\pi/18$ to $4\pi/9$ by the increment of $\pi/36$. Initial orientation $\theta(0)$ increases from $\pi/18$ to $17\pi/18$ by $\pi/18$ while the initial angular velocity is $\dot{\theta}(0)=0$. Parameters are set to $m=1$, $\kappa = 0.5$, and $\beta =0.5$.  }
\label{fig:bistability}
\end{figure*}

\begin{figure*}
\centering
 \includegraphics[scale = 1]{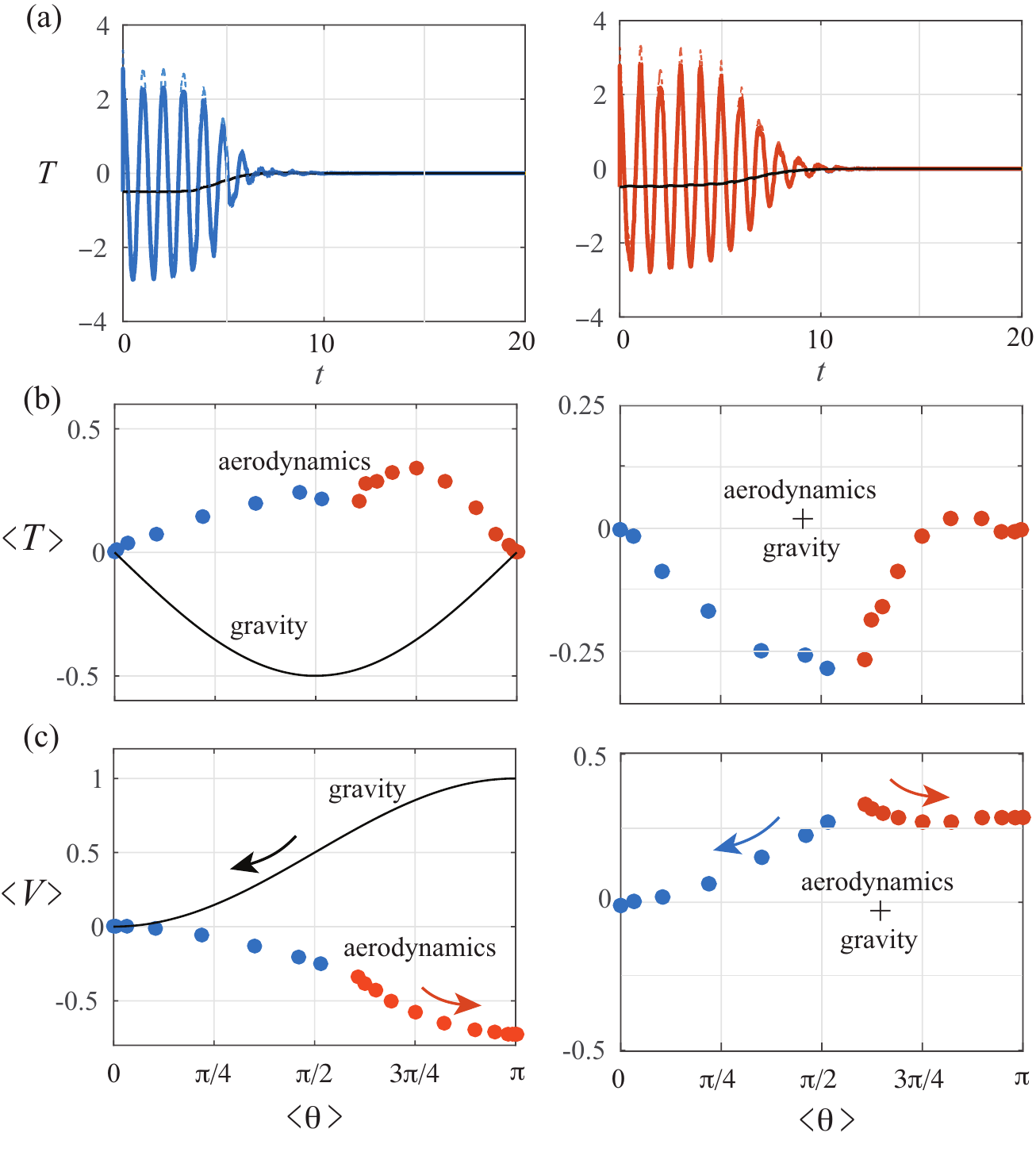}
\caption{\footnotesize (a) Torque as a function of time for the two trajectories highlighted in figure~\ref{fig:bistability}. (b) and (c) Time-averaged torque due to aerodynamics and gravity and corresponding rotational potential $V$ as a function of time-averaged orientation.}
\label{fig:potential}
\end{figure*}

\subsection{Basins of attraction of $\wedge$- and $\vee$-configurations} 

What is the value of the initial perturbation $\theta(0)$ beyond which the flyer stabilizes in the concave-up configuration? To answer this question, we vary the initial perturbation $\theta(0)$ from $0$ to $\pi$ by increments $\Delta \theta = \pi/18$, keeping track of the flyer's long-term behavior (concave-down or concave-up). The results are reported in figure~\ref{fig:bistability} for flow parameters $\beta = 0.5$ and $\kappa = 0.5$ and flyers of mass $m = 1$ and angle $\alpha$ ranging from $\pi/18$ to $4\pi/9$. The basin of attraction of the concave-up configuration increases as $\alpha$ increases, allowing for a stable concave-up configuration with perturbations from the upward direction as large as $\pi/2$. 
Figure~\ref{fig:bistability} also shows the time evolution of $\theta(t)$ given two initial conditions $\theta(0) = \pi/2$ and $\theta(0) = \pi/2 + \pi/18$ for a representative example of $\alpha = \pi/3$. The flyer converges to $\theta =0$ in one case and $\theta = \pi$ in the other.

For the classic inverted pendulum with $\beta$ and $\kappa$ satisfying the condition in~\eqref{eq:bistablependulum}, the limiting value $\theta_o$ of initial perturbations, averaged over the rapid vertical oscillations,  above which the pendulum is stable in the inverted configuration is given by~\citet[equation (9)]{Butikov2001},
\begin{equation}
\cos(\theta_o) = - \dfrac{1}{2\pi^2 \beta \kappa\cos\alpha}\left(\dfrac{2m_s}{3(m_s-m_f)}\right)^2.
\end{equation}
For $\kappa = \beta = 0.5$ as in figure~\ref{fig:bistability} and $m_s/m_f = 4$, as $\alpha$ increases from $\pi/18$ to $4\pi/9$, the angle $\theta_o$ marking the boundary of the basin of attraction between the downward stable and upward stable configurations increases from $0.55\pi$ to $0.87\pi$. Unlike the flyer, the basin of attraction of the inverted pendulum decreases as $\alpha$ increases.

\subsection{Effective rotational potential}

To elucidate the fluid mechanical basis of this bistability, we examine the total torques due to both aerodynamics and gravity for the two cases highlighted in figure~\ref{fig:bistability}. The torques  are shown in figure~\ref{fig:potential}(a) as a function of time. The two subplots  are practically indistinguishable because of the fast oscillations in the aerodynamic torque.
We therefore average the aerodynamic torque $T = T_1 +T_2$ and orientation $\theta$ over  one-period of background flow oscillations to obtain the `slow' quantities,
\begin{equation}
\langle T \rangle  = \int_t^{t+1}T(t')dt', \quad  \langle \theta \rangle  = \int_t^{t+1}\theta(t')dt',
\label{eq:slow}
\end{equation}
and we plot $\langle T \rangle$ versus $\langle \theta \rangle$ in figure~\ref{fig:potential}(b). The slow aerodynamic torque, shown in blue for $\theta(0) = \pi/2$ ($\wedge$-stable) and in red for $\theta(0) = \pi/2 + \pi/18$ ($\vee$-stable), is always positive, indicating that it is acting against gravity in both cases, albeit at slightly higher values in the latter. The torque due to gravity is shown in solid black line and the sum of both torques is shown in the right panel. As $\theta \to \pi$, the total torque in the $\vee$-stable case becomes positive; the aerodynamic torque overcomes the torque due to gravity. 

We define an effective potential function 
\begin{equation}
 V(t) = V(\theta(t)) = - \int_{\theta^\ast}^{\theta(t)}T(t)d\theta'(t),  
 \end{equation}
where $\theta^\ast = 0$ for the $\wedge$-stable case and $\theta^\ast = \pi$ 
for the $\vee$-stable case. Figure~\ref{fig:potential}(c) shows its slow evolution $\langle V \rangle$, defined according to~\eqref{eq:slow}, as a function of $\langle \theta \rangle$. The aerodynamic component of this potential counteracts the component due to gravity and dominates as $\theta$ approaches $\pi$ in the $\vee$-stable case, creating a `dip' in the potential around $\pi$. 

It is important to note that the results shown in figure~\ref{fig:potential}(c) do not represent the landscape of the potential function due to aerodynamic and gravitational torques. They rather correspond to a ``sampling" of this landscape by two particular trajectories.  To construct the aerodynamic potential, we fix the flyer at different angles $\theta$ ranging from $0$ to $\pi$ (no dynamics) and compute the aerodynamic forces and torque at each orientation as detailed next.

\subsection{Aerodynamic forces and torques and quasi-steady model}\label{sec:quasi-steady}

\begin{figure*}
\centering
 \includegraphics[scale = 1]{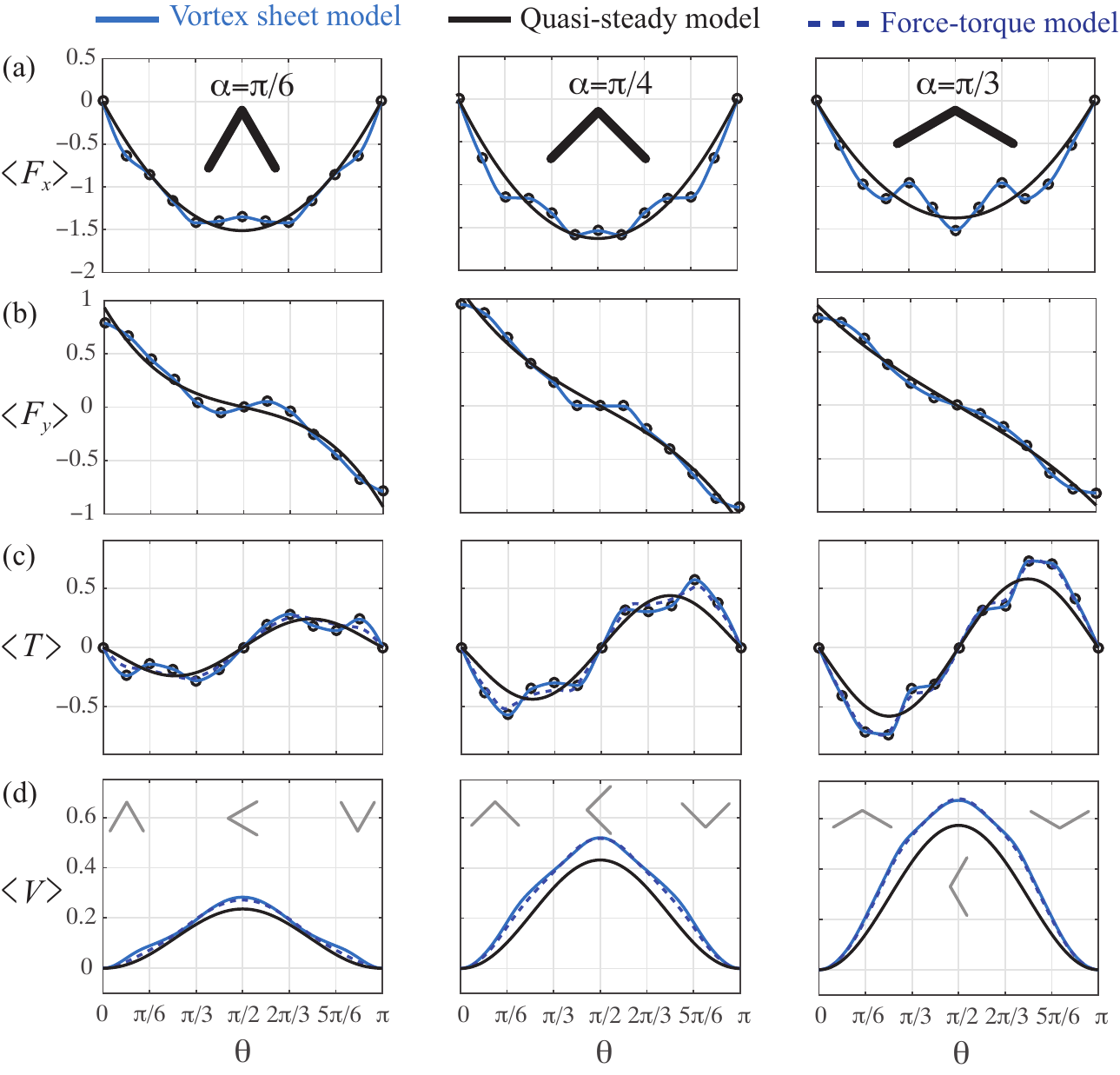}
\caption{\footnotesize  (a-c) Aerodynamic forces $\langle F_{x}\rangle$, $\langle F_{y}\rangle$, torque $\langle T \rangle$  averaged over one oscillation period as a function of $\theta$ based on the vortex sheet model (solid blue line) and the quasi-steady point force model (solid black line). (d)  Effective rotational potential $V$ as a function of $\theta$. Nominal parameter values are set to $m=1$ and $\kappa = \beta = 0.5$.
}
\label{fig:quasi}
\end{figure*}

 We fix the flyer at different angles $\theta$ ranging from $0$ to $\pi$ in a fluid oscillating with amplitude $\beta = 0.5 $ and acceleration $\kappa= 0.5$. At each orientation $\theta$, we compute the aerodynamic forces $\langle F_x \rangle$ and 
$\langle F_y \rangle$ based on the vortex sheet model (see~\eqref{eq:aero}) and averaged over fast flow oscillations. Results are shown in figure~\ref{fig:quasi}(a) and (b) for three flyers of half-opening angle $\alpha = \pi/6, \pi/4$ and $\pi/3$. Given the left-right symmetry of the flyer and the up-down symmetry of flow oscillations, $\langle F_x \rangle$ is symmetric about the horizontal axis $\theta = \pi/2$ while $\langle F_y \rangle$ is anti-symmetric. Importantly,  for $\theta<\pi/2$, $\langle F_y \rangle$ points in the opposite direction to gravity whereas for $\theta > \pi/2$, $\langle F_y \rangle$ reinforces gravity. 

We  postulate a quasi-steady point-force model that takes into account these symmetries as follows
\begin{equation}
\begin{split}
\label{eq:quasi}
\langle F_x \rangle &= A \theta( \pi - \theta ), \qquad \langle F_y \rangle = B \left(\dfrac{\pi}{2} - \theta  \right)^3 + C\left(\dfrac{\pi}{2} - \theta \right).
\end{split} 
\end{equation}
Here, the constant parameters $A$, $B$ and $C$ depend on the flyer's angle $\alpha$. The  values obtained from a least-square fit between the point-force model and the forces computed based on the vortex sheet model are listed in table \ref{table:coeffs}. 
The quasi-steady forces are superimposed on figures~\ref{fig:quasi}(a) and (b), showing good agreement with the vortex sheet model for all flyers.

We compute the aerodynamic torque about the flyer's point of attachment using~\eqref{eq:aero} and take its time-average $\langle T \rangle$ over the fast flow oscillations as in~\eqref{eq:slow}; see  figure~\ref{fig:quasi}(c).  The torque is anti-symmetric about the horizontal axis $\theta = \pi/2$: it is negative for $\theta < \pi/2$ (reinforcing gravity) and  positive for $\pi/2< \theta < \pi$. At first glance, this seems inconsistent with  figure~\ref{fig:potential}(c) where the aerodynamic torques act against gravity for all  $\langle \theta \rangle$ averaged over fast flow oscillations. {However, this discrepancy arises because} the plots in figure~\ref{fig:potential}(c) correspond to time-averaged values obtained from dynamic trajectories where the rotational momentum varies in time. In figure~\ref{fig:quasi}, the flyer is held fixed  in order to extract the inherent symmetries in the aerodynamic forces and torque induced by the oscillatory flow itself. {Further, note that this} analysis is consistent with the point-force model presented in~(\citet[figure 3]{Liu2012}) for the particular case $\alpha = \pi/6$. In~\citet{Liu2012}, the aerodynamic forces were postulated to act at the outer two edges of the flyer (at the sites of vortex emission) and  their directions and magnitudes were assumed to follow ad-hoc rules motivated by symmetry arguments. Based on these rules, the aerodynamic torque was computed about the flyer's center of mass.
Here, the aerodynamic forces and torque are computed exactly based on the vortex sheet model and the quasi-steady force model is built accordingly with no further assumptions. As such, it is applicable to flyers of any shape $\alpha$.

  \begin{table}
    \begin{center}
        \begin{tabular}{ c |  c c c c }
         $\alpha$ \qquad & \qquad A &  \qquad B & \qquad C  & \qquad D\\ \hline
         $\pi/6$ \qquad & \qquad -0.613 & \qquad 0.161 & \qquad 0.196 & \qquad 0.434 \\
         $\pi/4$ \qquad & \qquad -0.660 & \qquad 0.097 & \qquad 0.455 & \qquad 1.067 \\
         $\pi/3$ \qquad & \qquad -0.555 & \qquad 0.045 & \qquad 0.485 & \qquad 2.618 \\
        \hline
        \end{tabular}
    \caption{Coefficients of the quasi-steady model~\eqref{eq:quasi} and~\eqref{eq:torque} for the flyers shown in figure~\ref{fig:quasi}. 
    }
    \label{table:coeffs}
    \end{center}
\end{table}

Equations~\eqref{eq:aero} do not reflect the location of the aerodynamic center where the aerodynamic forces should be applied in order to produce an equivalent aerodynamic torque. To this end, we postulate that the force should act along the axis of symmetry of the flyer for all $\theta$ and we write
\begin{equation}
\begin{split}
\langle T  \rangle =  Dl\cos\alpha \left(\langle F_x \rangle  \cos\theta + \langle F_y  \rangle \sin\theta\right),
\label{eq:torque}
\end{split} 
\end{equation}
where $D$ is an unknown parameter that reflects the distance from the flyer's apex to the aerodynamic center. The values of $D$ listed in the last column of table \ref{table:coeffs} are obtained from a least-square fit between the values of $\langle T  \rangle$ computed directly from~\eqref{eq:aero} and those calculated from~\eqref{eq:torque} with forces computed from~\eqref{eq:aero}. For $\alpha = \pi/6$,  the aerodynamic center is close to the center of mass of the flyer ($D \approx 0.5$) as postulated in~\citet{Liu2012}. However, as $\alpha$ increases, $D$ also increases. For $\alpha = \pi/3$, $D$ is larger than five times the distance between the apex and {the center of mass}.

Lastly, we compute the rotational potential $\langle V \rangle$ due to aerodynamics such that $\langle T \rangle= - \partial  \langle V \rangle /\partial \theta$.  
Figure~\ref{fig:quasi}(d) shows three lines: the solid blue line is based on the vortex sheet model; the dashed blue line is based on the force-torque model in~\eqref{eq:torque} with forces obtained from the vortex sheet model;  the solid black line is based on~\eqref{eq:torque} and the quasi-steady model in~\eqref{eq:quasi}. The difference between the quasi-steady and vortex sheet models increases as the angle $\alpha$ of the flyer increases. For all $\alpha$, the aerodynamic potential is symmetric about $\pi/2$ and is characterized by two minima at $\theta =0$ and $\theta = \pi$. The potential wells around these minima are indistinguishable. This symmetry is broken in the presence of gravity. When the rotational potential $(m\beta/\kappa) \cos\alpha \cos \theta$ due to gravity is added, the well around $\theta = \pi$ becomes more shallow and disappears altogether when gravity is dominant.

In summary,  for $\theta < \pi/2$, as $\alpha$ increases from $\alpha = \pi/6$ to $\pi/3$, the $\wedge$-configuration gets more stable. At the same time,  the aerodynamic center gets pushed below the center of mass. Taken together, these two observations are consistent with the findings in~\cite{Liu2012} that top-heavy flyers are more stable. Meanwhile, 
For $\theta > \pi/2$, the same is true about the $\vee$-configuration. However, force calculations show that only the $\wedge$-configuration and perturbations smaller than $\pi/2$ produce aerodynamic forces that can potentially sustain the flyer's mass when released from the attachment point, as in~\citet{Weathers2010, Liu2012,Huang2015a, Huang2016}

\section{Results: elastic flyers}\label{sec:elastic}

\begin{figure*}
\centering
\includegraphics[width=0.75\linewidth]{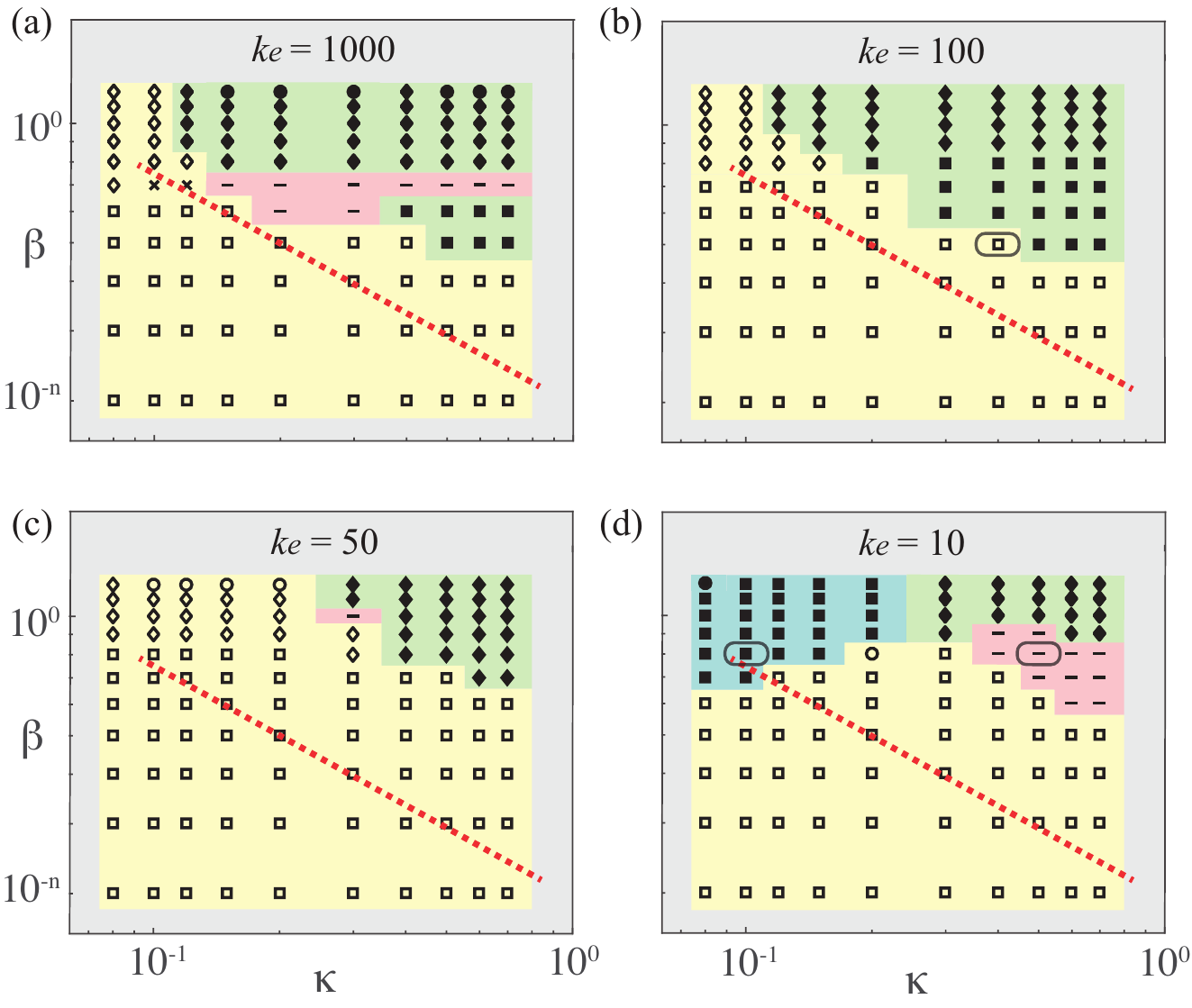}
\caption{\footnotesize Stable and bistable behavior of an elastic flyer mapped onto  the ($\kappa$,$\beta$) space for decreasing spring stiffness $k_e=1000, 100, 50, 10$. The mass and rest angle are set to $m=1$ and $\alpha_r = \pi/6$ as in the middle panel of figure~\ref{fig:paramspace}(b). 
}
\label{fig:paramspace_elastic}
\end{figure*}

\begin{figure*}
\centering
\includegraphics[width=0.75\linewidth]{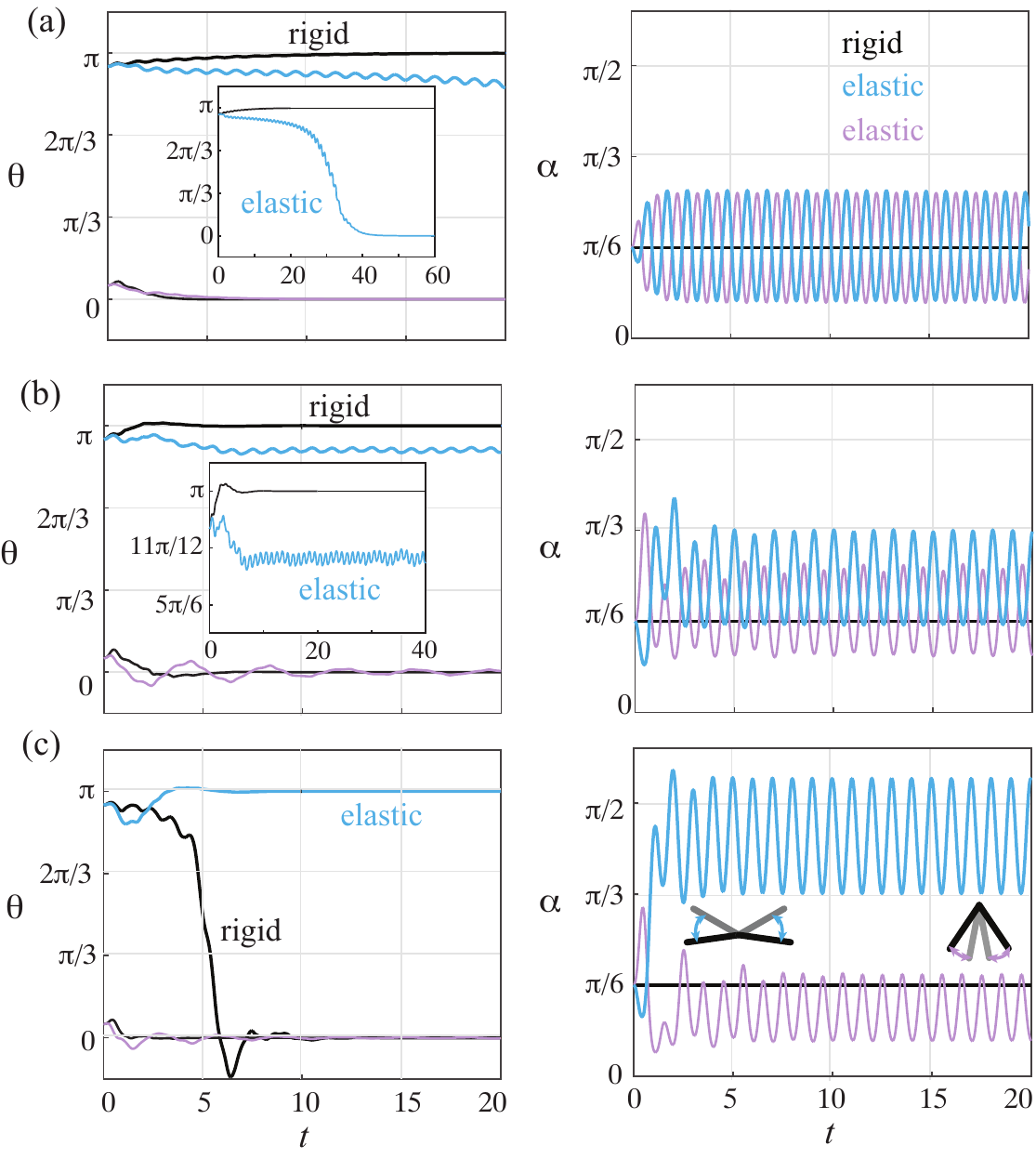}
\caption{\footnotesize  Elastic versus rigid flyers of mass  $m = 1$ and (rest) angle $\alpha_r = \pi/6$  for three sets of parameters highlighted in grey boxes in figure~\ref{fig:paramspace_elastic}. (a)  $\kappa = 0.4$, $\beta = 0.5$, $k_e = 100$,  (b) $\kappa = 0.5$, $\beta = 0.8$, $k_e = 10$, and (c) $\kappa = 0.1$, $\beta = 0.8$, $k_e = 10$. Initial conditions are $\theta(0) = \pi/18$ and $17\pi/18$, $\dot{\theta}(0)=0$, and $\alpha(0)-\alpha_r = \dot{\alpha}(0) = 0$.}
\label{fig:rigidelastic}
\end{figure*}

\begin{figure*}
\centering
\includegraphics[width=0.75\linewidth]{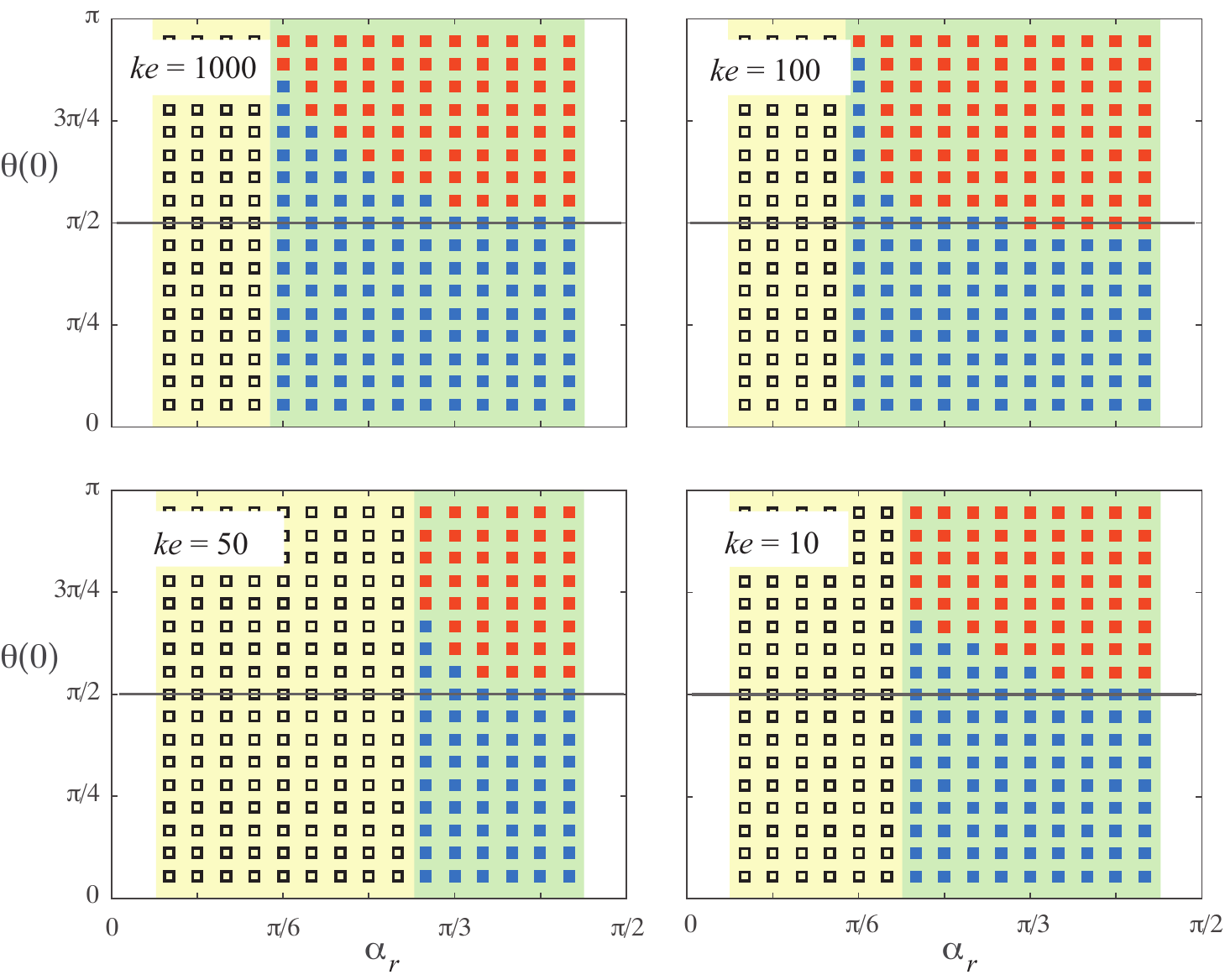}
\caption{\footnotesize Basins of attraction for the (blue) downward $\wedge$- and (red) upward $\vee$-stable configurations vary with the spring stiffness $k_e=1000, 100, 50, 10$ and the spring's rest angle $\alpha_r$.  Parameters are set to $\beta = 0.5$ and $\kappa = 0.5$ and $m = 1$.}
\label{fig:basin_elastic}
\end{figure*}

To examine the effect of flexibility on the flyer's response, we introduce a rotational spring at the apex between the two wings for a flyer of mass $m=1$. We fix the rest angle of the spring at $\alpha_r = \pi/6$ and consider four values of the stiffness coefficient: $k_e = 1000, 100, 50,$ and $10$. Smaller stiffness implies more compliant flyer. For infinitely large $k_e$, we recover the rigid flyer whose parameter space $(\beta,\kappa)$ is depicted in the middle panel of figure~\ref{fig:paramspace}(b). Here, we map the behavior of the elastic flyer onto the same parameter space  $(\beta,\kappa)$ for each value of $k_e$; see figure~\ref{fig:paramspace_elastic}. Similar to its rigid analog, the elastic flyer  exhibits stable and bistable behavior 
but the transition to bistable behavior is pushed up and to the right in the $(\kappa, \beta)$ plane.  {In other words,} the bistable region is smaller for $k_e = 1000$. The red line  in figure~\ref{fig:paramspace}(b) (middle panel)  marking the transition of the rigid flyer to bistability is overlaid onto the parameter space of the elastic flyer for ease of comparison. 

A new behavior is observed in flexible flyers at $k_e = 1000$. The new behavior is marked by `$-$' and highlighted in pink. It is characterized by the flyer being stable about an inclined orientation not equal to $\pi$. For $k_e=100$, the new behavior disappears and the bistable region increases slightly relative to that at $k_e = 1000$ but remains smaller than that of the rigid flyer.  As $k_e$ decreases to $50$, the new behavior reappears and the bistable region shrinks again, indicating that the size of the bistable region varies non-monotonically with $k_e$. In fact, it seems that $k_e = 100$ is optimal for maximizing the bistable region above the red line. Finally, for $k_e = 10$, the bistable behavior about inclined orientations reappears in the upper right region of $(\beta,\kappa)$ space. Importantly, bistable behavior appears 
 in the upper left corner at high values of $\beta$ and low values of $\kappa$ (region highlighted in blue). This new transition to bistability seems unique to highly flexible flyers, {and may be associated with the limit where gravitational and elastic forces are comparable, that is to say,  $O(m \beta/\kappa) \sim O(k_e)$ in (\ref{eq:eom})}.

To shed more light on the difference in behavior between the flexible flyer and its rigid analog, we show in figure~\ref{fig:rigidelastic} the time evolution of $\theta$ and $\alpha$  for three representative cases highlighted in grey boxes in figures~\ref{fig:paramspace_elastic}(b) and (d). 
Figure~\ref{fig:rigidelastic}(a) shows the flyer's orientation $\theta$ and flapping angle $\alpha$ about the rest angle $\alpha_r = \pi/6$ of the spring as functions of time for $\kappa = 0.4$, $\beta = 0.5$ and $k_e = 100$. Here, elasticity destabilizes the upward configuration.  

Figure~\ref{fig:rigidelastic}(b) shows the new behavior highlighted in pink in figure~\ref{fig:paramspace_elastic}. The parameter values are set to $\kappa = 0.5$, $\beta = 0.8$ and $k_e = 10$. The flyer stabilizes about an upward configuration around $\theta = 11\pi/12$ rather than $\pi$. The associated shape oscillations occur about a larger opening angle than the spring's rest angle.

Finally, figure~\ref{fig:rigidelastic}(c) shows the new transition to bistable behavior at $\kappa = 0.1$,  $\beta = 0.8$ and $k_e = 10$. The right panel of figure~\ref{fig:rigidelastic}(c) shows the flyer's flapping behavior. The inset schematics depict the range of flapping angles for the upward and downward stable trajectories. Because the flyer is compliant, it flaps about a much larger angle than the spring rest angle, thus increasing the effective opening angle of the flyer and the resulting aerodynamic torque. The flyer can therefore stabilize upward at much lower values of flow acceleration. However, in this flexible limit, the distinction between concave-up and concave-down is not very clear because the flyer exhibits both {types of} concavity over one oscillation cycle. 

In all three examples, the frequency of the flapping motion is equal to the frequency of the background flow, irrespective of initial conditions and parameter values. That is to say, the frequency of flapping $\alpha$ is slaved to aerodynamics rather than to the intrinsic natural frequency associated with the flyer's elasticity. We calculate the intrinsic natural frequency of the flyer as follows. We linearize~\eqref{eq:eom}, with aerodynamic torques set to zero, about the equilibrium configuration $(0,\alpha^\ast)$ of the `dry' system. To this end, $\alpha^\ast$ is given by
\begin{equation}
\sin\alpha^* = \frac{k_e\kappa}{m\beta}(\alpha_r-\alpha^*), 
\end{equation}
and $\alpha_r -{m\beta}/{k_e\kappa} \le \alpha^* \le \alpha_r$. The linear equations are 
\begin{equation}
\begin{split}
\delta\ddot{\theta} + (\frac{\beta}{\kappa}\cos\alpha^* ) \delta\theta  = 0 , \qquad  
\delta\ddot{\alpha} + (\frac{k_e}{m}+\frac{\beta}{\kappa}\cos\alpha^*)\delta\alpha = 0.
\end{split}
\end{equation}
The first equation leads to the rotational natural frequency of the classic pendulum.  The natural frequency $f_{n}^\alpha$ of shape oscillations follows from the second equation,
\begin{equation}
f_{n}^\alpha = \frac{1}{2\pi}\sqrt{\frac{k_e}{m}+\frac{\beta}{\kappa}\cos\alpha^*}.
\end{equation}
For $k_e=10$, the natural frequency $f_{n}^\alpha$ is about 1/2.

Lastly, we examine the effect of elasticity on the `basin of attraction' of the vertically-upward configuration. Figure~\ref{fig:basin_elastic} shows that, in comparison with the rigid flyer in figure~\ref{fig:bistability}, the introduction of a stiff spring $k_e=1000$ has a small effect on the basin of attraction of $\theta = \pi$. As $k_e$ decreases, this basin seems to increase  and it is maximum at $k_e = 100$. As $k_e$ decreases further ($k_e = 50$), the region of bistable behavior decreases but not the basin of attraction. Finally, for $k_e  = 10$, both the region of bistable behavior and the basin of attraction of $\theta=\pi$ increase, certainly due to an increase in the effective opening angle of the compliant flyer.

\section{\label{sec:level5}Conclusions}

The main contributions of this work can be summarized as follows.
\begin{itemize}
\item[] (i) We considered the rotational stability of a $\wedge$-flyer of half-opening angle $\alpha$ attached at its apex  and free to rotate in a vertically oscillating flow.  The flyer is always stable about the downward $\wedge$-configuration. Depending on flow parameters, aerodynamics can stabilize the flyer about the upward $\vee$-configuration. We analyzed the transition from stable to bistable behavior as a function of dimensionless flow amplitude and acceleration.
\item[]  (ii) We compared this aerodynamically-induced transition to bistability with the inertia-induced transition of a classic pendulum undergoing vertical base oscillations.
In both cases, the transition happens for oscillation amplitudes $\beta$ and accelerations $\kappa$ satisfying $\beta/\kappa^a > b$,  with $-1<a<0$ for the flyer and $a=-1$ for the pendulum. The transition to bistable behavior depends on $\alpha$. {For the flyer, increasing $\alpha$ facilitates this transition and enlarges the basin of attraction for the upward configuration.  In contrast, for the inverted pendulum, larger $\alpha$ hinders this transition to bistable behavior.}

\item[] (iii) Using the vortex-sheet model, we computed the aerodynamic forces, averaged over fast flow oscillations, as a function of the flyer orientation $\theta$. We found that the horizontal force is symmetric and the vertical force is anti-symmetric about up-down reflections. These symmetries exist for all angles $\alpha$ and can be easily traced back to the left-right symmetry of the flyer and up-down symmetry of the oscillating background flow. 

Based on these computations, we postulated a quasi-steady point force model whose coefficients depend on the flyer's angle $\alpha$.
\item[] (iv) We computed the aerodynamic torque, averaged over fast flow oscillations, and calculated the rotational potential  associated with the slowly-varying torque. The aerodynamic potential is symmetric about up-down reflections; it is characterized by two minima at the $\wedge$-  and $\vee$-configurations irrespective of $\alpha$. The two wells are deeper for larger $\alpha$, indicating more stable behavior for flyers with wider opening angles, as noted in~\citet{Huang2015a}. Gravity breaks this symmetry in favor of the $\wedge$-configuration. 
\item[] (v) Lastly, we considered the effect of flexibility on the flyer's behavior by introducing a rotational spring at its apex. The flyer flaps passively due to the background flow oscillations. Flexibility diminishes upward stability in stiff flyers, but a new transition to upward stability is observed in compliant flyers. 
\end{itemize}

Our force calculations show that due to up-down asymmetry, $\wedge$-flyers can use aerodynamic forces to support their weight only when $\theta < \pi/2$, in agreement with~\citet{Weathers2010, Liu2012,Huang2015a, Huang2016}. Further, our results suggest that stable $\wedge$-configurations can be maintained by manipulating either the opening angle or stiffness of the flyer. 
These findings will guide the development of future research aimed at understanding the rotational stability of biological and bio-inspired flyers.
Insects use flight muscles attached at the base of the wings to flap~\citep{Pringle2003}.
Insect wings and flight muscles are thought to be stiff~\citep{Ellington1985} but organisms can modulate their muscle stiffness \citep{Feldman2009}.
 It is therefore plausible that, by manipulating the stiffness of their flight muscle, insects can maintain stability in the face of environmental disturbances.

\paragraph*{Acknowledgment.} The work of Y.H. and E.K. is supported by the National Science Foundation (NSF) through the grants NSF CMMI 13-63404 and NSF CBET 15-12192 and by the Army Research Office (ARO) through the grant W911NF-16-1-0074.

\bibliographystyle{jfm}
\bibliography{reference}

\end{document}